\documentclass[%
 reprint,
 amsmath,amssymb,
 aps,
floatfix
]{revtex4-2}

\usepackage{graphicx}
\usepackage{dcolumn}
\usepackage{bm}

\usepackage{ulem} 

\usepackage{xcolor}

\newcommand{\blue}[1]{\textcolor{blue!0!black}{#1}}

\newcommand{\evdel}[1]{\langle #1 \rangle}

\usepackage{subcaption} 
\usepackage{svg}
\usepackage{amsthm}
\usepackage{amsmath}
\usepackage{blindtext}
\usepackage{todonotes}
\usepackage{comment}

\begin{document}

\title{Effective Topology}
\title{Route-preserving network reduction for on-demand ride-pooling models}

\title{Emergence of cyclic routes from on-demand ride-pooling}
\title{Cyclic routes approximate  on-demand ride-pooling}
\title{Dynamics-preserving effective networks for distance-based routing processes}
\title{Dynamics-preserving network reductions for \blue{ride-pooling paths}}

\author{K. Stiller}%
 \affiliation{Potsdam Institute for Climate Impact Research,Telegrafenberg A 31, Potsdam, Germany}
 \affiliation{Technical University Berlin, Straße des 17. Juni 135, Berlin 10623, Germany}

\author{N. Molkenthin}
\email{nora.molkenthin@pik-potsdam.de}
 \affiliation{Potsdam Institute for Climate Impact Research,Telegrafenberg A 31, Potsdam, Germany}
\date{\today}

\begin{abstract} 
Reducing the complexity of \blue{ride-pooling paths} is a central challenge in systems with distributed demand. Here we show that such dynamics admit an exact coarse-grained representation: for a broad class of routing algorithms whose decisions depend only on path lengths, the full network can be reduced to an effective network of active nodes weighted by shortest-path distances without altering the resulting trajectories, up to stochastic degeneracy breaking.
The reduction therefore defines an equivalence class of network representations generating identical path dynamics.
We further demonstrate that for globally optimizing dispatchers this equivalence is systematically violated through degeneracy amplification, yet remains quantitatively accurate beyond the exactly solvable regime.
Our results identify when spatial structure can be integrated out without loss of dynamical fidelity, providing a general framework for the analysis of interacting path processes.

\end{abstract}

\maketitle

\section{Introduction}
Many dynamical processes on networks can be understood as ensembles of paths generated online under local or global constraints. Examples range from interacting random walks \cite{barbier2022self} and polymers \cite{rubin1965random} to computer science applications \cite{xia2019random}. In such systems, the microscopic geometry of the network often contains a large number of degrees of freedom that do not directly enter the decision-making dynamics. Identifying when and how these degrees of freedom can be integrated out without altering the resulting dynamics is a central problem in coarse-graining of non-equilibrium systems.

In contrast to classical models such as random or self-avoiding walks \cite{noh2004random}, many real-world path-based processes are neither memoryless nor strictly repulsive with respect to their own history. Instead, paths interact through soft constraints, shared resources, or evolving cost landscapes, leading to partially self-avoiding and mutually interacting trajectories.
Online ride-pooling systems \cite{zwick2022ride, schwieterman2018sharing,tang2017data} provide a \blue{practically relevant} realization of such interacting path dynamics. The pooling algorithms generate trajectories on a spatial network in response to successive origin–destination requests, depending on the planned paths of the entire fleet \cite{creutzig2024shared,zwick2022review}. 

Although already widely used, the dynamics of the passenger transport sector in general and shared pooled mobility in particular is not yet fully understood. This is largely due to the enormous complexity of the problem arising from a large number of variables and parameters, such as request locations and timing, the street network, vehicle interactions, and routing algorithms. Additionally, pooling becomes harder as the number of origins, destinations and alternative paths between them increases. To combat this problem many real-world ride-pooling providers \cite{agatz2011dynamic,spieser2014toward, zwick2021agent,harmann2023development} as well as theoretical approaches \cite{lotze2022dynamic, lotze2024taming} use stop-pooling (also called meeting points or virtual stops) to reduce the number of times vehicles have to stop to pick up or drop off customers.
Effectively stop-pooling re-assigns trip requests to a subset of nodes, designated as pooling stops (in the following called \textit{active nodes}). These are typically denser than traditional line-based public transport stops but sparser than a door-to-door service.
Pooled stops create a spiky demand pattern, where demand to and from the active nodes is high and demand to and from all other nodes (in the following called \textit{passive nodes}) zero.

Here we combine the spiky demand pattern and the street network topology into one much smaller \textit{effective network}, in which all passive nodes are eliminated while preserving shortest-path distances between the dynamically relevant active nodes. \blue{This greatly reduces system complexity and thus conceptually simplifies the analysis.} We demonstrate that for a broad class of distance-based dispatchers, this reduction preserves the path dynamics, yielding path ensembles and performance measures identical up to degeneracy-breaking decisions.

 \begin{figure*}[t]
\centering
\subfloat[Original network with five active nodes]{
\minipage{0.25\textwidth}
  \includegraphics[width=\linewidth]{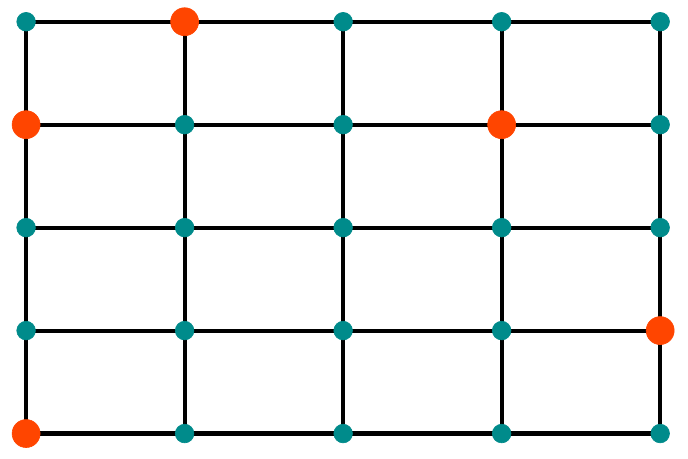}
\endminipage}\hfill
\subfloat[Complete Request Graph]{
\minipage{0.25\textwidth}
  \includegraphics[width=\linewidth]{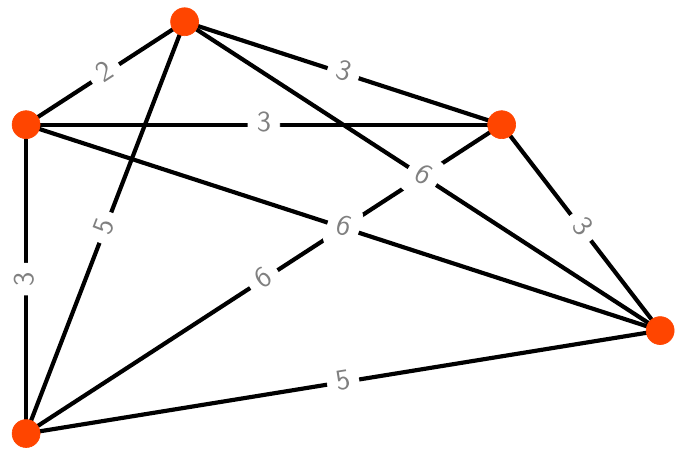}
\endminipage}\hfill
\subfloat[Minimal Effective Topology]{
\minipage{0.25\textwidth}%
  \includegraphics[width=\linewidth]{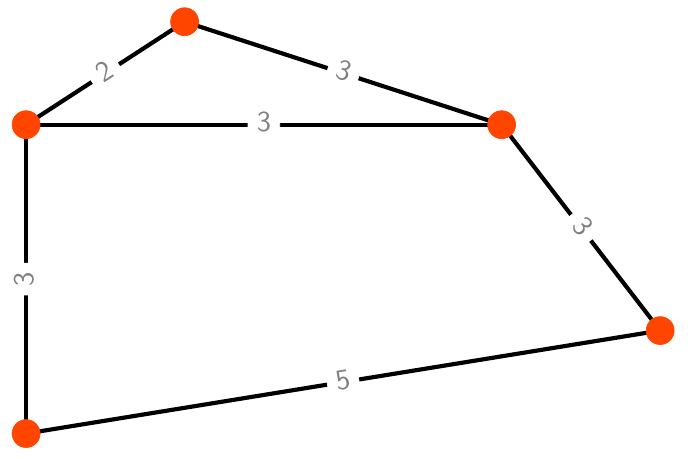}
\endminipage}
\caption[Mapping of an artificial street network with 5 active nodes to the corresponding complete request graph and minimal effective topology]{
The original network a) has active nodes (red) and passive nodes (green). Networks in the effective topology equivalence class only contain active nodes but preserve all distances between them (b,c). This can be achieved by constructing a complete graph b) with edge weights accounting for the average shortest path length between each pair of active nodes of the minimal effective topology c) which only contains links between pairs of active nodes, which have no shortest path going through another active node.}
\label{fig:effectivetopologyclass}

\end{figure*}

\section{Theoretical derivation of effective topology}

Online ride-pooling routes are the result of a dynamic process, which continuously generates requests as timed origin-destination pairs in a Poisson process with an average time $\Delta t$ between incoming requests. An online dispatcher algorithm typically inserts requests optimizing a cost function. Cost functions may focus on customer travel time \cite{molkenthin2020scaling}, driven vehicle distance \cite{jung2024ridepy,schmaus4899318shared} or combinations of the two, even including vehicle re-balancing or allowing changes between vehicles \cite{namdarpour2024non}. Despite their differences, all of these dispatcher algorithms base their optimization on a) the planned routes of the vehicle fleet and b) the distances/travel times along the streets of the street network.

Mathematically, these are expressed as the fleet\blue{'s currently scheduled stop sequences together with the weighted} street network.
The fleet stop sequence of the ride-pooling fleet at time $t$ is defined as
\begin{equation}
    \mathcal{F}_t=\{F^{(1)}_t,F^{(2)}_t,...,F^{(B)}_t\},
\end{equation}
where $F^{(1)}_t,F^{(2)}_t,...,F^{(B)}_t$ are the stop sequences of the individual vehicles
\begin{align}
    F^{(1)}_t,&=\{f^{(1)}_1,f^{(1)}_2,...,f^{(1)}_{n_1}\}, \nonumber \\
    F^{(2)}_t,&=\{f^{(2)}_1,f^{(2)}_2,...,f^{(2)}_{n_2}\}, \nonumber \\
   &\vdots \nonumber \\
    F^{(B)}_t,&=\{f^{(B)}_1,f^{(B)}_2,...,f^{(B)}_{n_B}\}
    \label{eq.F}
\end{align}
of lengths $n_i$. Note that $f^{(i)}_1$ is the current location of vehicle $i$ at time $t$ (even though the index $t$ has been dropped for the individual stops in Eq.\ref{eq.F} for readability).

Most online ride pooling algorithms follow a common pattern: a new request $(u,v)$ is drawn from the request distribution $D(G)$ on the street network $G$. Then a cost function is evaluated, assigning a cost to every possible insertion choice and selecting the one with the lowest cost. The request $(u,v)$ is then inserted according to the insertion choice associated with the lowest cost.

Here we define the effective topology as the equivalence class of network topologies $\mathcal{G}_e$ from the set of all networks $\mathcal{G}$, which replicate the same \blue{fleet stop sequence  $\mathcal{F}_t$} as the original system at all times, while using a uniform request distribution $U(G)$
on all of its nodes.
\begin{equation}
    \mathcal{G}_e=\{G'\in \mathcal{G}: \mathcal{F}_t^{OR}(D(G))=\mathcal{F}_t^{ET}(U(G')) \;\; \forall t\},
\end{equation}
\blue{where $\mathcal{F}_t^{OR}$ is the fleet stop sequence in the original system and $\mathcal{F}_t^{ET}$ is the fleet stop sequence in the effective topology.}

Let us assume a binary request distribution $D(G)$, which is zero in some nodes (\textit{passive nodes}) and uniform across the remaining nodes (\textit{active nodes}). 

\blue{By design of the generic ride-pooling algorithm described above all networks in $\mathcal{G}_e$ must contain all active nodes and none of the passive nodes, as only active nodes will appear in the fleet stop sequence in equal quantities. As the ride-pooling algorithm uses the distances between nodes in the cost-function, these distances must be matched to the path lengths between the nodes along the original network in $\mathcal{G}_e$. Thus the active nodes are connected by links weighted by the shortest-paths distance along the original network.
The equivalence class always contains the complete graph $G_c$ of active nodes (metric closure of the subgraph of active nodes), as well as any network that can be constructed by removing links from this complete request graph without changing the shortest path length between any pair of nodes (See Fig.~\ref{fig:effectivetopologyclass}).} We call the Graph with the fewest links in $\mathcal{G}_e$ the minimal effective topology $G_m$. The minimal effective topology contains a link between two active nodes if and only if there is no shortest path between them that goes through another active node.
Thus, $\mathcal{G}_e$ contains all networks of all active nodes, which preserves the distance matrix of the active nodes in the original network.

To \blue{show that this construction preserves} the dynamics on the original network $G$, consider two consecutive stops $(f_i,f_{i+1})$ on a vehicle route. Let
\begin{equation}
    \mathcal{I}_G(f_i,f_{i+1})
\end{equation}
denote the set of active nodes that can be inserted between these stops \blue{on the original network} under the dispatcher’s cost function, and define $\mathcal{I}_{G_e}(f_i,f_{i+1})$ analogously \blue{on an effective network $G_e \in \mathcal{G}_e$}.
\begin{figure*}
    \centering
    \includegraphics[width=0.8\textwidth]{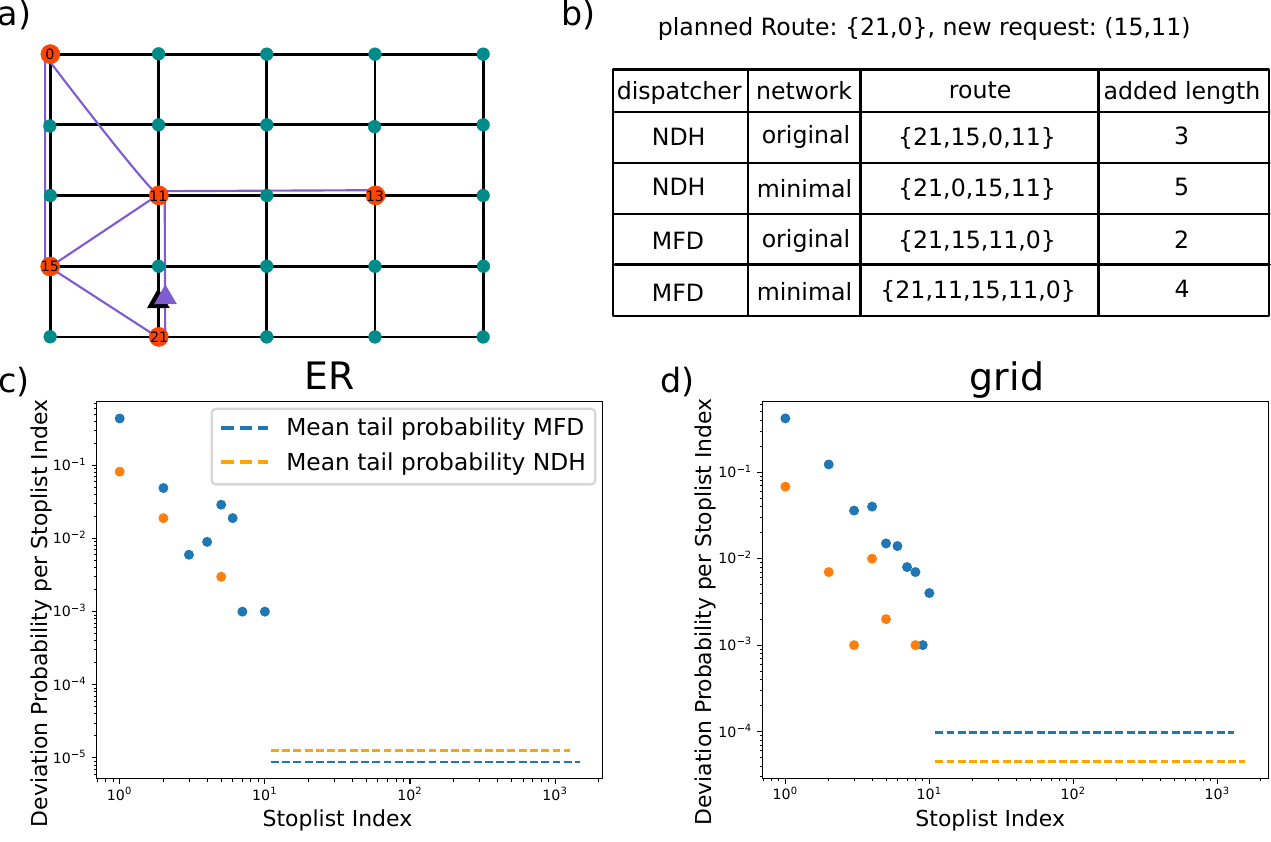}
    \caption{\blue{Geometric degeneracy breaking affecting the insertion decision. a) 5x5 grid network with 5 active nodes (red) together with the effective network (purple). b) Table showing insertions in a planned route from node 21 to node 0. A request comes in just after the vehicle has left node 21, leading to four different insertions, depending on the combination of dispatcher algorithm and network. c) Probability of an insertion difference by cumulative stop list index on Erdos-Renyi network with 100 nodes. The cumulative stop list is the final stop list without removal of served nodes. Most insertions occur in the first 10 steps. d) Probability of an insertion difference by cumulative stop list index on square grid network with 100 nodes.}}
    \label{fig:illustration3}
\end{figure*}

By construction, the effective network preserves all pairwise shortest-path distances between active nodes,
\begin{equation}
    d_G(u,v)=d_{G_e}(u,v) \text{     for all active nodes $u$,$v$.}
\end{equation}
For insertion heuristics that evaluate candidates exclusively through these distances — for instance by minimizing additional path length \blue{resulting from the insertion of a node $s$ between $u$ and $v$}
\begin{equation}
    \Delta(u,s,v)=d(u,s)+d(s,v)-d(u,v),
\end{equation}
possibly subject to a detour tolerance — admissibility and cost ranking depend only on the distance matrix between active nodes. Consequently,
\begin{equation}
    \mathcal{I}_G(f_i,f_{i+1})=\mathcal{I}_{G_e}(f_i,f_{i+1})
\end{equation}

and the minimizing insertion choice coincides on both networks. Since this holds for every insertion step, the resulting fleet stop lists evolve identically on $G$ and $G_e$. In this sense, the effective network is \blue{exactly} dynamically equivalent to the original network for all insertion heuristics whose decisions depend solely on inter-stop distances.
\begin{figure*}
    \centering
    \includegraphics[width=\textwidth]{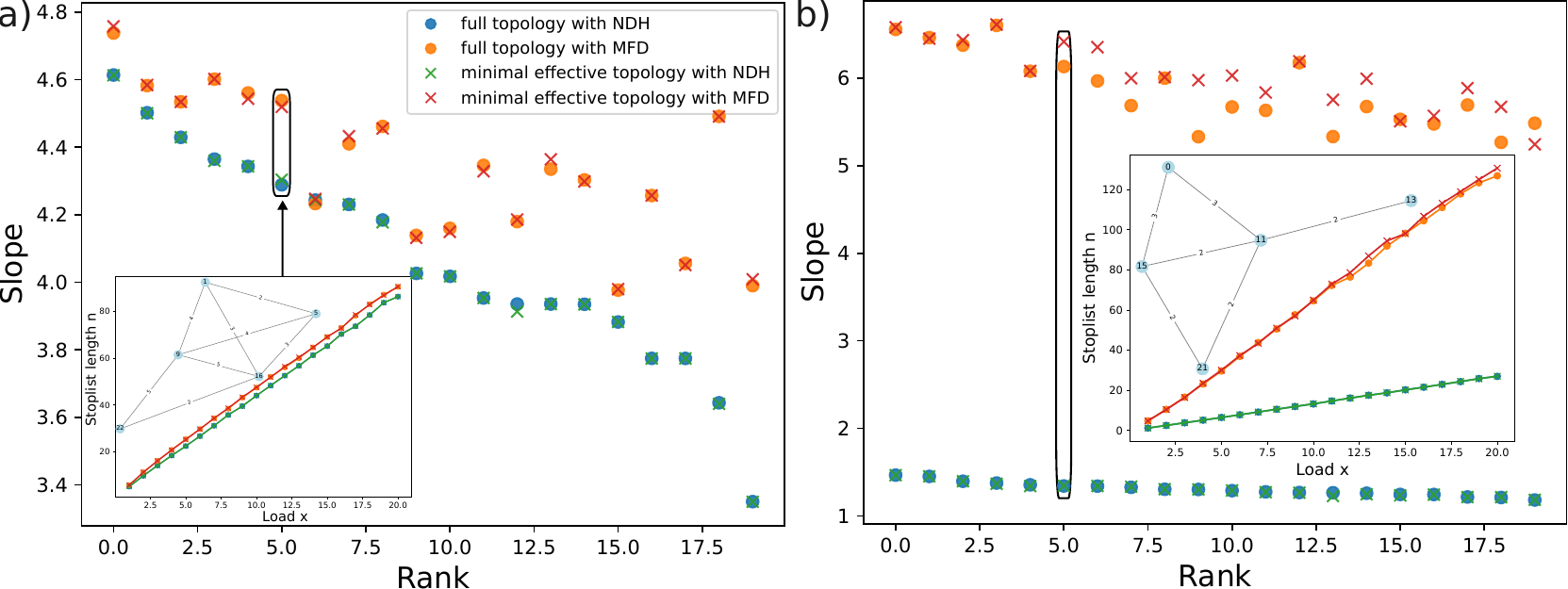}
    \caption{
    Simulations on 20 full and minimal effective topologies result in identical slopes for both dispatchers. Simulations are done with a maximum waiting time of 20, a maximum delay factor of 5 and a seat capacity of 80 simulated for 20000 requests and displayed in descending order of \blue{NDH} slope. a) For a fleet size of 1. The resulting slopes of the two dispatchers are \blue{distinct yet} similar\blue{, since fleet distance and individual arrival time are very closely related. However, the MFD dispatcher has the flexibility to take shortcuts, delaying individual requests}. b) For a fleet size of 50. The difference between the two dispatchers becomes apparent, the difference between full and minimal effective topology increases slightly for the brute force dispatcher.}
    \label{fig:slope}
\end{figure*}

Deviations can arise only if the dispatcher’s decision rule depends on geometric information beyond the stop list — for example, on the instantaneous vehicle position along an edge or on non–shortest-path structural features of the underlying graph. Fig.~\ref{fig:illustration3} shows a concrete example \blue{for the network shown in Fig.~\ref{fig:illustration3} a), with the minimal effective network (purple edges) and the vehicle position shown as black and purple triangles. The table in Fig.~\ref{fig:illustration3} b) illustrates an insertion of a new request just after the vehicle has left 21, heading to 0 with four different results. However, simulations in Fig.~\ref{fig:illustration3} c and d) show that this type of deviation is rare. Here a single vehicle was simulated over 1000 realizations of 1000 requests at a normalized system load of $x=20$, as defined in \cite{molkenthin2020scaling} as:
\begin{equation}
    x = \frac{\evdel{l}_{\text{\tiny{D(G)}}}}{vB} \, \lambda \,,
    \label{eq:dimensionlessLoad}
\end{equation}
where $\evdel{l}_{\text{\tiny{D(G)}}}$ is the average requested shortest path length for the network $G$ with the request distribution $D(G)$, $\lambda$ is the request frequency of the Poisson process, $B=1$ is the fleet size and $v$ is the vehicle speed, which we set to $v=1$ for simplicity.}
\blue{Fig.~\ref{fig:illustration3} c) shows the probability of a first deviation over the stop sequence index on an Erdos-Renyi network with 100 nodes, of which 5 are active. Fig.~\ref{fig:illustration3} d) shows the same measure on a 100 node grid network. While the probability is still considerable at the first or second insertion, it drops off quickly to a very low baseline level once the planned route has reached equilibrium length. This is because deviations in insertions between original and effective topology can only happen if two planned consecutive stops are not direct neighbours on the effective network. This situation becomes rare at high request rates.}

In other words, the deviation stems from a difference in the timing of the decision, which path is taken. Such geometric slack may create additional admissible insertions in the original network that are absent in the reduced representation. In the absence of such state-dependent geometric effects, however, the fleet dynamics are fully determined by the preserved distance structure, and the effective network yields identical routing sequences.

\section{Effective Topology in simulations}
\blue{We have shown in the previous section that effective networks retain the exact path dynamics up to geometric ambiguity. In this section we explore the practically relevant differences between full and effective networks across sets of five randomly selected active nodes on a $5 \times 5$ grid.
Here we use the average length of the resulting equilibrium stop sequences $\evdel{n}$ as a simplified measure of ride pooling efficiency or average passenger travel time.
average stop sequence length $\evdel{n}$ is defined as the average over time, vehicle and realization, of the number of scheduled stops in the stop lists $\mathcal{F}_t$.}

As established in \cite{molkenthin2020scaling}, the length of the average stop sequence is approximately proportional to the system load:
\begin{equation}
    \evdel{n}=\frac{2 \evdel{t_w}+\evdel{t_d}}{\tau} x \approx S x,
    \label{eq.n}
\end{equation}
where $\evdel{t_w}$ is the average waiting time and $\evdel{t_d}$ the average time customers spend aboard the vehicle and $\tau=\frac{\evdel{l}_{\text{\tiny{D(G)}}}}{v}$ is the characteristic time of the system, or the average direct driving time for a request.

The simulations presented here use two different dispatcher algorithms: the No-Detour-Heuristic dispatcher (NDH) and the Minimal-Fleet-Distance dispatcher (MFD).
The NDH dispatcher introduced in \cite{schmaus2026emergence} minimizes the arrival time of the new request, without any additional detours to already scheduled trips. To this end all positions on and after the routes are evaluated for arrival time, selecting the one with the quickest delivery of the request.
The MFD dispatcher, introduced in \cite{jung2024ridepy} tests all combinations of insertions of pick-up and drop-off, which are consistent with the constraints of maximum waiting time $t^{\text{max}}_{w}$, maximum delay factor \blue{$d=t^{\text{max}}_{d}/t^{\text{direct}}$} and vehicle capacity. Here, we chose relatively high values of $t^{\text{max}}_{w}=20$, $d=5$ and vehicle capacity \blue{$C=80$}, in order eliminate rejections and keep the two dispatchers comparable. The algorithm uses the total distance added to the route as a cost function, selecting the insertion, for which the added distance is the lowest.

In Fig.~\ref{fig:slope} (insets) $\evdel{n}$ is shown to be approximately linear with system load for both dispatcher algorithms, implying that $\evdel{t_w}$ and $\evdel{t_d}$ are approximately constant with $x$. Fig.~\ref{fig:slope} shows the slopes of 20 random node configurations for both dispatchers on the full topology as well as the minimal effective topology for fleet sizes of a) 1 bus or b) 50 buses. \blue{In Fig.~\ref{fig:slope} a) the slopes of full and effective topology agree well for both dispatchers. The MFD dispatcher has slightly steeper slopes, because NDH minimizes the passenger travel time directly ($\evdel{t_w}+\evdel{t_d}$), while MFD minimizes the route length. }In Fig.~\ref{fig:slope} b) the slope for the MFD dispatcher is much higher as a result of the rejection-minimizing parameter choices. The values for full and minimal effective topology remain remarkably close in both cases. This indicates that the minimal effective topology is a valid simplification for networks with a restricted set of active nodes, regardless of load or dispatcher algorithm. 

While both insertion rules depend on pairwise shortest-path distances \blue{together with the geometric locations of vehicles on the current edge,} the global fleet-distance minimization explores a substantially larger combinatorial decision space. The resulting increase in degeneracy of optimal insertions amplifies the effect of stochastic tie-breaking, \blue{especially for larger fleets}. This leads to more frequent divergence between realizations on the original and reduced networks, despite identical distance structures and thus larger differences between orange circles and red crosses in Fig.~\ref{fig:slope} b).

\section{Discussion and Conclusion}
We introduced the effective topology as a distance-preserving sparsification of on-demand ride-pooling networks with active request nodes and passive transit nodes. The construction reduces network size while preserving all pairwise shortest-path distances between active nodes. For dispatchers whose insertion decisions depend exclusively on fleet state and inter-stop distances, this reduction reproduces the fleet dynamics exactly, \blue{with stochastic degeneracy breaking resulting from the geometric information of vehicle positions along the current edges.} This establishes an equivalence class of network representations that generate identical routing trajectories under any distance-determined heuristic.
Simulations with NDH and MFD dispatchers demonstrate that the effective network remains an accurate approximation, \blue{when edge position is included}, with deviations controlled by the system load and \blue{fleet size}. 

\blue{Despite the substantial reduction in system complexity, simulation times were not significantly reduced. However, this simplification deepens our understanding of the relationship of the underlying street network and the demand distribution as two influences on the fleet stop list, that can effectively be used interchangeably, at least for binary demand distributions.} This mathematically justifies simplifications already commonly made in practical implementations of on-demand ride-pooling services based on stop-pooling, meeting points or virtual stops.

The derivation relies only on the fleet state and the network distance matrix, implying that the reduction applies to any routing dynamics whose decisions depend solely on these quantities, or equivalently on travel times. More broadly, the results identify conditions under which spatial network geometry can be integrated out. This perspective suggests that analogous distance-preserving reductions may be applicable in other networked systems with active and passive nodes, including supply chains\cite{yamano2023supply}, information spreading with influencer–follower dynamics \cite{milli2018active, chung2019susceptible}, and reduced models of power grids \cite{hartmann2024synchronized}.

\bibliographystyle{unsrt}

\bibliography{ecobuslit}
\end{document}